\documentclass[12pt,a4paper]{article}
\pdfoutput=1
\usepackage[text={16.5cm,23cm},centering]{geometry}
\usepackage{amsmath,amssymb}

\newcommand{\Ocal}{\mathcal{O}}
\newcommand{\Zb}{\mathbb{Z}}
\newcommand{\fvev}[1]{\left\langle #1\right\rangle_{\mathrm{free}}}
\newcommand{\dfvev}[1]{\left\langle #1\right\rangle_{\mathrm{defect}}}
\newcommand{\dOPE}[1]{#1}
\newcommand{\eqfree}{\overset{\mathrm{free}}{=}}

\begin{document}
\begin{center}
  \begin{flushright}
    OU-HET 901
  \end{flushright}
  \vspace{8ex}
  {\Large \bfseries \boldmath The $\epsilon$-expansion of the codimension two twist defect from conformal field theory}\\
  \vspace{4ex}
  {\Large Satoshi Yamaguchi}\\
  \vspace{2ex}
  {\itshape Department of Physics, Graduate School of Science, 
  \\
  Osaka University, Toyonaka, Osaka 560-0043, Japan}\\
  \vspace{1ex}
  \texttt{yamaguch@het.phys.sci.osaka-u.ac.jp}\\
  \begin{abstract}
    We apply the framework of Rychkov-Tan \cite{Rychkov:2015naa} to the codimension two twist defect at the Wilson-Fisher fixed point in $4-\epsilon$ dimensions. We obtain the scaling dimensions of the operators on the defect up to the lowest nontrivial order in the $\epsilon$-expansion without using Feynman diagram computation. Our results agree with the known results.
  \end{abstract}
\end{center}

\vspace{4ex}

Recently, there has been remarkable progress in the conformal bootstrap program after the seminal work \cite{Rattazzi:2008pe}.
In particular, the numerical results \cite{ElShowk:2012ht,El-Showk:2014dwa,Gliozzi:2013ysa,Gliozzi:2014jsa,Kos:2014bka,Simmons-Duffin:2015qma,Kos:2016ysd} suggest that 3-dimensional Ising CFT is completely determined by the conformal symmetry and certain reasonable ansatz.
Now it is an important open question to explain why and how the conformal symmetry constrains the theory so strongly.

The analytic approach to this question started by Rychkov and Tan \cite{Rychkov:2015naa} is a promising one.  They study $(4-\epsilon)$-dimensional Wilson-Fisher (WF) theory by employing a set of reasonable axioms, and determine the leading correction of the scaling dimensions of a class of operators without using Feynman diagram calculation.  This direction is further developed in \cite{Basu:2015gpa,Ghosh:2015opa,Raju:2015fza,Sen:2015doa,Nii:2016lpa}.

In this letter we study the twist defect introduced by Bill\'o, Caselle, Gaiotto, Gliozzi, Meineri, and Pellegrini \cite{Billo:2013jda}, and further studied in \cite{Gaiotto:2013nva}.  This defect is codimension 2 and defined by monodromy around the defect.  This theory contains a class of primary operators localized on the defect, denoted by $\psi_s,\ (s\in\Zb+1/2)$.

We apply the framework of \cite{Rychkov:2015naa} to this twist defect in this letter.  We obtain the leading order correction to the scaling dimensions $\Delta_s$ of $\psi_s$ without using Feynman diagram calculation. Our result agrees with the known result obtained by Feynman diagram calculation in \cite{Gaiotto:2013nva}.  We conclude that the  framework of \cite{Rychkov:2015naa} works even if a defect is inserted.


Let us first consider the single real free scalar field theory in 4 dimensions as preparation for the WF theory.
We use the notations of \cite{Rychkov:2015naa} for bulk operators in this letter. The normalization of the free scalar field $\varphi(x)$ is determined by the two-point function as
\begin{align*}
\fvev{\varphi(x_1)\varphi(x_2)}=\frac{1}{|x_1-x_2|^{2}}.
\end{align*}

We introduce the ``twist defect'' as introduced and studied from various points of view in \cite{Billo:2013jda,Gaiotto:2013nva}.  This twist defect is defined by monodromy as follows.
Let $x^{\mu},\ \mu=1,\dots,4$ be the coordinates of 4-dimensional space. We insert the codimension two defect at $x^1=x^2=0$.  There is monodromy around this defect given by
\begin{align*}
  \varphi(r\cos(\theta+2\pi),r\sin(\theta+2\pi),x^3,x^4)
  =-\varphi(r\cos\theta,r\sin\theta,x^3,x^4).
\end{align*}
This monodromy defines the theory of this defect.

The operators on this defect include primary operators $\psi_s,\ (s\in \Zb+1/2)$  \cite{Billo:2013jda,Gaiotto:2013nva}. Each $\psi_s$ has the smallest scaling dimensions among spin $s$ operators.  The relation between $\varphi$ and $\psi_s$ is expressed by the bulk-defect operator product expansion (OPE).  When the operator $\varphi$ is inserted at $x=(r\cos\theta,r\sin\theta,0,0)$ close to the defect, it is expanded by the operators localized on the defect.  This bulk-defect OPE of $\varphi(x)$ is found by making use of the two-point function of $\varphi$ \cite{Gaiotto:2013nva}.  The bulk-defect OPE of $\varphi(x)$ is given by
\begin{align}
  \dOPE{\varphi(x)}\eqfree \cdots+e^{is\theta}r^{|s|} \psi_s(0)+\cdots.
\label{V1d}
\end{align}
The notation $\eqfree$ means the equality in the 4-dimensional free theory.
The terms $\cdots$ include operators of other spins than $s$ or terms with higher powers of $r$.
Although the phases of the OPE coefficients are not fixed by the two-point function, they are absorbed by the definition of $\psi_s$.
In this letter we fix this phase by the OPE \eqref{V1d}.  Another result of \cite{Gaiotto:2013nva} used in this letter is the expectation value of $\varphi^2(x)$ in the presence of the defect:
\begin{align*}
  \dfvev{\varphi^2(x)}\eqfree -\frac{1}{8r^2}.
\end{align*}
From these results and Wick's theorem, we obtain the following bulk-defect OPE of $\varphi^3$.
\begin{align}
  \dOPE{\varphi^3(x)}\eqfree \cdots-\frac{3}{8}e^{is\theta}r^{|s|-2} \psi_s(0)+\cdots.
\label{V3d}
\end{align}

Next let us turn to the WF theory in $4-\epsilon$ dimensions. We use the framework of \cite{Rychkov:2015naa}, and employ basically the same axioms.
The difference is that our axioms in this letter include the twist defect.
\paragraph{Axiom I}
\label{par:Axiom I}
The WF theory is conformally invariant. The twist defect is invariant under the conformal transformations which fix the defect geometrically.
\paragraph{Axiom II}
\label{par:Axiom II}
For an arbitrary operator $\Ocal_{\mathrm{free}}$ in the 4-dimensional free theory (on the defect) there is an operator $\Ocal_{\mathrm{WF}}$ in the WF theory (on the defect) such that
\begin{align*}
\lim_{\epsilon \to 0} \Ocal_{\mathrm{WF}}=\Ocal_{\mathrm{free}}.
\end{align*}

Axiom II implies that there exists a local operator $V_n$ in the WF theory which satisfies $\lim_{\epsilon\to 0}V_n=\varphi^n$.  Axiom II also implies the existence of the counterpart of $\psi_s$ in the WF theory. We use the same notation $\psi_s$ for the WF theory and the 4-dimensional free theory.  The third axiom states the relation between $V_1$ and $V_3$.

\paragraph{Axiom III}
\label{par:Axiom III}
$V_3$ is not a primary operator but a descendant of $V_1$ i.e.
\begin{align*}
\Box V_1=\alpha V_3,
\end{align*}
where $\alpha$ is a constant and $\Box$ is the Laplacian.

We need the following two facts obtained in \cite{Rychkov:2015naa} starting from the axioms. The constant $\alpha$ in Axiom III is given by
\begin{align}
  \alpha=\frac{2}{9}\epsilon+O(\epsilon^2).\label{alpha}
\end{align}
The scaling dimension of $V_1$ denoted by $\Delta_1$ is given by
\begin{align}
  \Delta_1=1-\frac12 \epsilon + O(\epsilon^2).\label{Delta1}
\end{align}

After the above preparation, the scaling dimensions $\Delta_s$ of $\psi_s$ are obtained as follows.  Axiom I implies the existence of the bulk-defect OPE.  The bulk-defect OPE of $V_1(x)$ contains the leading spin $s$ term, which is fixed by the scale symmetry, as
\begin{align}
\dOPE{V_1(x)}=\cdots + C_{1s}\frac{e^{is\theta}}{r^{\Delta_1-\Delta_s}} \psi_s(0)+\cdots,\label{WFV1d}
\end{align}
where $C_{1s}$ is a constant.
From Axiom III, the bulk-defect OPE of $V_3$ is written as
\begin{align}
\dOPE{V_3(x)}
=&\frac{1}{\alpha}\dOPE{\Box V_1(x)}\nonumber\\
=&\cdots + \frac{1}{\alpha}C_{1s}\Box\frac{e^{is\theta}}{r^{\Delta_1-\Delta_s}} \psi_s(0)+\cdots,\nonumber\\
=&\cdots + \frac{-s^2+(\Delta_1-\Delta_s)^2}{\alpha}C_{1s}\frac{e^{is\theta}}{r^{\Delta_1+2-\Delta_s}} \psi_s(0)+\cdots.\label{WFV3d}
\end{align}
By comparing \eqref{WFV1d} and \eqref{V1d} we obtain $\lim_{\epsilon\to 0}C_{1s}=1$.
Similarly by comparing \eqref{WFV3d} and \eqref{V3d} we obtain
\begin{align}
  -\frac{3}{8}
  =\frac{-s^2+(\Delta_1-\Delta_s)^2}{\alpha}C_{1s}+O(\epsilon).
  \label{C3s}
\end{align}
Notice that $-s^2+(\Delta_1-\Delta_s)^2$ must be $O(\epsilon)$ since $\alpha=\frac{9}{2}\epsilon=O(\epsilon)$ (see eq.~\eqref{alpha}).  Actually, multiplying both sides of \eqref{C3s} by $\alpha$ and using $C_{1s}=1+O(\epsilon)$, we obtain
\begin{align}
  -s^2+(\Delta_1-\Delta_s)^2=-\frac{3}{8}\alpha + O(\epsilon^2).
\label{C3s2}
\end{align}
We substitute $\alpha$ and $\Delta_1$ of eq.~\eqref{alpha} and eq.~\eqref{Delta1}  into eq.~\eqref{C3s2}, and obtain
\begin{align}
\Delta_s=|s|+1+\left(-\frac12-\frac{1}{24|s|}\right)\epsilon+O(\epsilon^2).
\label{Deltas}
\end{align}
The expression \eqref{Deltas} is the final result of this letter.
It agrees with the result in \cite{Gaiotto:2013nva}.

\subsection*{Acknowledgement}
This work was supported in part by JSPS KAKENHI Grant Number 15K05054.

\bibliographystyle{utphys}
\bibliography{refs}
\end{document}